\documentclass[a4paper]{spie}  

\usepackage[]{graphicx}

\title{Optomechanical detection of weak forces} 

\author{David Vitali, Stefano Mancini and Paolo Tombesi
\skiplinehalf
INFM and Dept. of Physics, University of Camerino, I-62032 Camerino Italy
}

\begin{document} 
  \maketitle 

\begin{abstract}
Optomechanical systems are often used for the measurement of weak
forces. Feedback loops can be used in these systems for
achieving noise reduction.
Here we show that even though
feedback is not able to improve the signal to noise ratio of the device
in stationary conditions, it is possible to design a nonstationary
strategy able to improve the sensitivity. 
\end{abstract}



\section{INTRODUCTION}
\label{sect:intro}  

Optomechanical devices are used in high sensitivity
measurements, as the 
interferometric detection of gravitational waves 
\cite{GRAV}, and 
in atomic force microscopes \cite{AFM}. 
Up to now, the major limitation to the implementation
of sensitive optical measurements
is given by thermal noise \cite{HADJAR}.
Some years ago it has been proposed \cite{MVTPRL} to reduce thermal noise
by means of a feedback loop based on homodyning the light 
reflected by the oscillator, playing the role of a cavity mirror.
This proposal has been then experimentally 
realized~\cite{HEIPRL,PINARD,ATTO} 
using the ``cold damping'' technique \cite{COLDD}, 
which is physically analogous to that proposed in Ref.~\cite{MVTPRL} and which
amounts to applying a viscous feedback force to the oscillating mirror.
In these experiments, the viscous force is provided
by the radiation pressure of another laser beam, intensity-modulated by the
time derivative of the homodyne signal.

Both the scheme of Ref.~\cite{MVTPRL} and the cold 
damping scheme of Refs.~\cite{HEIPRL,PINARD,ATTO} 
cool the mirror by overdamping it, 
thereby strongly decreasing its mechanical susceptibility at resonance.
As a consequence, the oscillator does not resonantly respond to the 
thermal noise, yielding
in this way an almost complete suppression of the resonance peak in the 
noise power spectrum, which is equivalent to cooling.
However, the two feedback schemes cannot be directly applied 
to improve the detection
of weak forces. In fact
the strong reduction of the mechanical susceptibility at resonance means that
the mirror does not respond not only to the noise but also to the signal.
This means that the signal to noise ratio (SNR) of the device
in stationary conditions is actually never improved 
\cite{LETTER,PRALONG}.
Despite that, here we show how it is possible to
design a {\em nonstationary} strategy
able to significantly increase the SNR for the detection
of {\em impulsive} classical forces acting on the oscillator.
This may be useful 
for microelectromechanical systems, where the search for quantum 
effects in mechanical systems is very active \cite{MEMS}, 
as well as for the detection of gravitational waves \cite{GRAV}.

\section{The model}

\begin{figure}
   \begin{center}
   \begin{tabular}{c}
   \includegraphics[height=4cm]{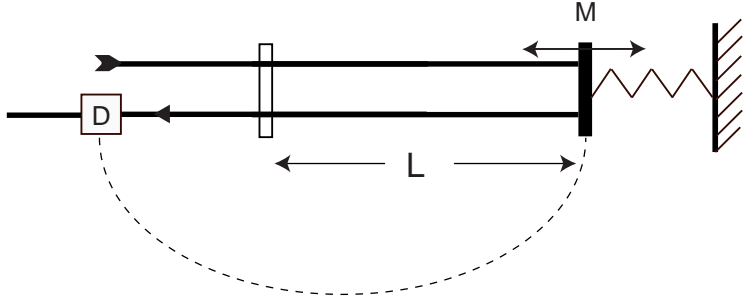}
   \end{tabular}
   \end{center}
   \caption[coolosa_fig1] 
   { \label{coolosa_fig1} 
 Schematic description of a linear Fabry-Perot cavity
with the end oscillating mirror M.
The equilibrium cavity length is $L$.
A cavity mode is driven by an input laser beam.
The output field is subjected to homodyne detection (D).
The signal is then fed back to the mirror motion (dashed line).}
\end{figure}

We shall consider a simple example of optomechanical system,
a Fabry-Perot cavity with a movable end mirror
(see Fig.~\ref{coolosa_fig1} for a schematic description).
The optomechanical coupling between the mirror and
the cavity field is realized by the radiation pressure. The 
electromagnetic field exerts a force which is 
proportional to the intensity of the field, which, at the same time, 
is phase-shifted by an amount 
proportional to 
the mirror displacement from its equilibrium position.
The mirror motion is the result of the 
excitation of many vibrational modes, including internal acoustic modes.
However here we shall study the {\em spectral} detection of weak forces
acting on the mirror and in this case one can focus
on a single mechanical mode only, by considering
a detection bandwidth including 
a single mechanical resonance peak
\cite{hadjar2}.
Optomechanical devices are able to reach the sensitivity limits imposed
by quantum mechanics \cite{PRALONG}
and therefore we describe the mechanical mode as a single {\em quantum} 
harmonic oscillator with mass $m$ and frequency 
$\omega_{m}$. 

When $\omega_{m}$ is much smaller 
than the cavity free spectral range, 
one can focus on one cavity mode only, because
photon scattering into other modes can be neglected
\cite{LAW}. Moreover, in this regime,
the generation of photons due to the Casimir effect,
and also retardation and Doppler effects are completely negligible
\cite{HAMI}.
The system is then described by the following two-mode 
Hamiltonian \cite{HAMI}
\begin{equation}
	H=\hbar \omega_{c}b^{\dagger}b + 
    \hbar\omega_m\left(P^{2}+
	Q^{2}\right) 
	-2\hbar G
	b^{\dagger}b Q -2\hbar f(t) Q + 
	i\hbar E\left(b^{\dagger}e^{-i\omega_{0}t}-b
	e^{i\omega_{0}t}\right) \,, 
	\label{HINI}
\end{equation}
where $b$ is the cavity mode annihilation operator with optical
frequency $\omega_{c}$, $f(t)$ is the classical force to be detected,
and $E$ describes the coherent input field
with frequency $\omega_{0}\sim \omega_{c}$
driving the cavity.
The quantity $E$ is related to the input laser 
power $\wp$ by $E=\sqrt{\wp\gamma_{c}/\hbar \omega_{0}}$,
being $\gamma_c$ the photon decay rate.
Moreover, $Q$ and $P$ are the dimensionless position and momentum
operator of the movable mirror. It is $\left[Q,P\right]=i/2$, 
and $G=(\omega_c/L)\sqrt{\hbar/2m\omega_m}$ represents the 
optomechanical coupling constant, with $L$
the equilibrium cavity length.

The dynamics of an optomechanical system is also influenced
by the dissipative interaction
with external degrees of freedom. The cavity mode is damped 
due to the
photon leakage through the mirrors which couple the cavity 
mode with the continuum of the outside electromagnetic modes. 
For simplicity 
we assume that the movable mirror has perfect reflectivity and that
transmission takes place through the fixed mirror only.
The mechanical mode undergoes Brownian motion caused by the 
uncontrolled coupling with other internal and external modes at 
thermal equilibrium. We shall neglect in our treatment
all the technical noise sources:  
we assume that the driving laser is stabilized
in intensity and frequency and we neglect the electronic noise
in the detection circuit. Including these 
supplementary noise sources is however quite straightforward (see for 
example \cite{KURT}). 
Moreover recent experiments have shown
that classical laser noise can be made negligible in the relevant 
frequency range \cite{HADJAR,TITTO}.

The dynamics of the system
can be described by the following set of coupled quantum 
Langevin equations (QLE)
(in the interaction picture with respect to 
$\hbar \omega_{0}b^{\dagger} b$) 
\begin{eqnarray}
\dot{Q}(t) &=& \omega_m P(t) \,, 
\label{QLENL1}\\
\dot{P}(t) &=& -\omega_{m} Q(t) + {\cal W}(t) +f(t) -  
 {\gamma_m} P(t)   
 +G b^{\dagger}(t) b(t) \,,
\label{QLENL2}\\
\dot{b}(t) &=& - \left(i \omega_{c} - i \omega_{0} +
\frac{\gamma_{c}}{2}\right) b(t) + 2 i 
G Q(t) b(t) + E +
\sqrt{\gamma_{c}}b_{in}(t)\,, 
\label{QLENL3}
\end{eqnarray}
where $b_{in}(t)$ is the input noise operator \cite{GAR} 
associated with the vacuum 
fluctuations of the continuum of modes 
outside the cavity, having the 
following correlation functions 
\begin{eqnarray}
&& \langle b_{in}(t)b_{in}(t') \rangle = \langle 
b_{in}^{\dagger}(t)b_{in}(t') \rangle
= 0 \,,
\label{INCOR1}\\
&& \langle b_{in}(t)b_{in}^{\dagger}(t') \rangle = \delta(t-t') \,.
\label{INCOR2} 
\end{eqnarray}
Furthermore, ${\cal W}(t)$ is the quantum Langevin force acting on 
the mirror, with the following correlation function \cite{VICTOR},
\begin{equation}
\langle {\cal W}(t) {\cal W}(t^\prime) \rangle=
\frac{\gamma_m}{4 \pi \omega_m} 
\int_{-\varpi}^{\varpi} d\omega \;
  \omega e^{-i\omega (t-t')}\left[ \coth\left(\frac{\hbar\omega}{2 k_B T} 
  \right)+1\right]
\label{BROWCOR}
\end{equation}
with $T$ the bath temperature, $\gamma_m$ the mechanical 
decay rate, $k_B$ the 
Boltzmann constant, and $\varpi$ the frequency cutoff of the reservoir
spectrum. 

In standard applications the driving field is
very intense so that the system is characterized by a
semiclassical steady state with 
the internal cavity mode in a coherent
state $|\beta\rangle $, and a new equilibrium position for the 
mirror, displaced by $G| \beta |^{2}/\omega_m$ 
with respect to that with no driving field.
The steady state amplitude $\beta $ is given by the solution
of the classical nonlinear equation 
$\beta=E/(\gamma_{c}/2+i \omega_{c} - i \omega_{0}
	+2 i G^2/\omega_{m}
        |\beta|^{2})$.
In this case, the dynamics is well described 
by linearizing the QLE 
(\ref{QLENL1})-(\ref{QLENL3}) around the steady state. Redefining
with $Q(t)$ and $b(t)$ the quantum 
fluctuations around the classical steady state, 
introducing the field phase 
$Y(t)=i\left(b^{\dagger}(t)-b(t)\right)/2$ and field amplitude 
$X(t)=\left(b(t)+b^{\dagger}(t)\right)/2$,  and choosing
the resulting cavity mode detuning 
$\Delta = \omega_{c} - \omega_{0} +
	2G^2/\omega_m \beta^{2}=0$
(by properly tuning the driving
field frequency $\omega_{0}$), the 
linearized QLEs can be rewritten as
\begin{eqnarray}
\dot{Q}(t) &=& \omega_m P(t) \,, 
\label{QLE2L1}\\
\dot{P}(t) &=& -\omega_{m} Q(t) -  
  {\gamma_m} P(t)   +2 G\beta X(t) + {\cal W}(t) +f(t)\,,
\label{QLE2L2}\\
\dot{Y}(t) &=&  -\frac{\gamma_{c}}{2} Y(t)
+2 G \beta Q(t)+ \frac{\sqrt{\gamma_{c}}}{2} Y_{in}(t) \,,
\label{QLE2L3} \\
\dot{X}(t) &=&  -\frac{\gamma_{c}}{2} X(t)
+ \frac{\sqrt{\gamma_{c}}}{2} X_{in}(t) \,,
\label{QLE2L4}
\end{eqnarray}  
where we have introduced the phase input noise 
$Y_{in}(t)=i\left(b_{in}^{\dagger}(t)-b_{in}(t)\right)$ and the 
amplitude input noise $X_{in}(t)=b_{in}^{\dagger}(t)+b_{in}(t)$.

\section{Position measurement and feedback}

As it is shown by Eq.~(\ref{QLE2L3}),
when the driving and the cavity fields are resonant, the dynamics is 
simpler because only the phase quadrature $Y(t)$ is 
affected by the mirror position fluctuations $Q(t)$,
while the amplitude quadrature $X(t)$ is not.  
Therefore the mechanical motion
of the mirror can be detected by monitoring
the phase quadrature $Y(t)$. 
The mirror position measurement is commonly performed in the 
large cavity bandwidth limit $\gamma_{c} \gg G \beta $, $\omega_m$,
when the 
cavity mode dynamics adiabatically follows that of the movable mirror
and it can be eliminated, that is, from Eq.~(\ref{QLE2L3}),
\begin{equation}
Y(t) \simeq \frac{4G \beta}{\gamma_{c}}Q(t) 
+\frac{Y_{in}(t)}{\sqrt{\gamma_{c}}},
\label{adiab}
\end{equation}
and $X(t) \simeq X_{in}(t)/\sqrt{\gamma_{c}}$ from Eq.~(\ref{QLE2L4}).
The experimentally detected quantity is the output homodyne photocurrent
\cite{HOW,GTV,homo}
\begin{equation}\label{BOUNDARY}
Y_{out}(t)=2\eta \sqrt{\gamma_{c}}Y(t)-\sqrt{\eta}Y_{in}^{\eta}(t)\,,
\end{equation}
where $\eta$ is the detection efficiency and $Y_{in}^{\eta}(t)$
is a generalized phase input noise, coinciding with the
input noise $Y_{in}(t)$ in the case of perfect detection $\eta 
=1$, and taking into account the additional noise due to the 
inefficient detection in the general case $\eta < 1$ \cite{GTV}.
This generalized phase input noise can be written in terms 
of a generalized input noise $b_{\eta}(t)$ as
$Y_{in}^{\eta}(t)=i\left[b_{\eta}^{\dagger}(t)-b_{\eta}(t)\right]$. 
The quantum noise $b_{\eta}(t)$ is correlated with the 
input noise $b_{in}(t)$ and it is characterized by the following 
correlation functions \cite{GTV}
\begin{eqnarray}
&& \langle b_{\eta}(t)b_{\eta}(t') \rangle = \langle 
b_{\eta}^{\dagger}(t)b_{\eta}(t') \rangle
= 0 \,,
\label{INCORETA1}\\
&& \langle b_{\eta}(t)b_{\eta}^{\dagger}(t') \rangle = \delta(t-t') 
\,,
\label{INCORETA2} \\
&& \langle b_{in}(t)b_{\eta}^{\dagger}(t')
\rangle = \langle b_{\eta}(t)b_{in}^{\dagger}(t')
\rangle = \sqrt{\eta}\delta(t-t').
\label{INCORETA3}  
\end{eqnarray}
The output of the homodyne measurement may be used to devise a
phase-sensitive feedback loop to control the dynamics of the mirror,
as in the original proposal\cite{MVTPRL}, 
or in cold damping schemes\cite{HEIPRL,PINARD,ATTO,COLDD}.
Let us now see how these two feedback schemes modify the 
quantum dynamics of the mirror.

In the scheme of Ref.\cite{MVTPRL}, the feedback loop induces a 
continuous position shift controlled by the output homodyne 
photocurrent $Y_{out}(t)$.
This effect of feedback manifests itself in an additional term
in the QLE for a generic operator ${\cal O}(t)$ given by
\begin{equation}\label{DOSTRA}
\dot{{\cal O}}_{fb}(t)=
i\frac{\sqrt{\gamma_{c}}}{\eta}
\int_{0}^{t} dt'G_{mf}(t')Y_{out}(t-t')
\left[g_{mf}P(t),{\cal O}(t)\right]\,,
\end{equation}
where $G_{mf}(t)$ is the feedback transfer function, and $g_{mf}$ is a 
feedback gain factor.  The implementation of this scheme is nontrivial
because it is equivalent to add a feedback interaction linear in the mirror 
momentum, as it could be obtained with a charged mirror
in a homogeneous magnetic field. For this reason here we shall refer to it
as ``momentum feedback'' (see, however, the recent parametric 
cooling scheme demonstrated in Ref.~\cite{ATTO}, showing some 
similarity with the feedback scheme of Ref.~\cite{MVTPRL}).

Feedback is characterized by a delay time which is essentially determined 
by the electronics and is always much smaller than the typical 
timescale of the mirror dynamics. It is therefore common to consider
the zero delay-time limit $G_{mf}(t) \sim \delta(t)$. For 
linearized systems, the limit
can be taken directly in Eq.~(\ref{DOSTRA}) \cite{PRALONG}, 
so to get the following QLE in the presence of feedback 
\begin{eqnarray}
\dot{Q}(t)&=&\omega_{m}P(t)
+g_{mf}\gamma_{c}Y(t)-\frac{g_{mf}}{2}\sqrt{
\frac{\gamma_c}{\eta}}Y_{in}^{\eta}(t)\,,
\label{QFBEQ1}\\
\dot{P}(t) &=& -\omega_{m} Q(t) -  
  {\gamma_m} P(t)   +2 G\beta X(t) + {\cal W}(t)+f(t)\,,
\label{QFBEQ2}\\
\dot{Y}(t) &=&  -\frac{\gamma_{c}}{2} Y(t)
+2 G \beta Q(t)+ \frac{\sqrt{\gamma_{c}}}{2} Y_{in}(t) \,,
\label{QFBEQ3} \\
\dot{X}(t) &=&  -\frac{\gamma_{c}}{2} X(t)
+ \frac{\sqrt{\gamma_{c}}}{2} X_{in}(t) \,,
\label{QFBEQ4}
\end{eqnarray} 
where we have used Eq.~(\ref{BOUNDARY}).
After the adiabatic elimination of the radiation mode (see 
Eq.~(\ref{adiab})), and introducing the rescaled, dimensionless, input 
power of the driving laser 
$ \zeta =  16 G^{2}\beta ^{2}/\gamma_m\gamma_{c}=
64G^{2}\wp/\hbar \omega_{0}\gamma_{m}\gamma_{c}^{2}$, 
and the rescaled feedback gain
$g_1 = -4G\beta g_{mf}/\gamma_{m}$, 
the above equations reduce to
\begin{eqnarray}
\dot{Q}(t)&=&\omega_{m}P(t)
-\gamma_m g_{1} Q(t)-\sqrt{\frac{\gamma_{m}}{\zeta}}g_{1}
Y_{in}(t)+\sqrt{\frac{\gamma_{m}}{\eta \zeta}}\frac{g_{1}}{2}
Y_{in}^{\eta}(t)\,,
\label{QEQ1}\\
\dot{P}(t)&=&-\omega_{m}Q(t)
-\gamma_{m}P(t)+\frac{1}{2}\sqrt{\gamma_{m}\zeta}
X_{in}(t)+{\cal W}(t)+f(t)\,.
\label{QEQ2}
\end{eqnarray}
This treatment explicitly includes the limitations due to the 
quantum efficiency of the detection, but neglects other
possible technical imperfections of the feedback loop, as for
example the electronic noise of the feedback loop, whose effects have 
been discussed in \cite{PINARD}.

Cold damping techniques
have been applied in classical electromechanical systems
for many years \cite{COLDD}, 
and only recently they have been proposed to improve
cooling and sensitivity at the quantum level \cite{GRAS}.
This technique is based on the application of a negative derivative
feedback, which increases the damping of the system without 
correspondingly increasing the thermal noise \cite{GRAS}.
This technique has been succesfully applied to an optomechanical 
system composed of a high-finesse cavity with a movable mirror 
in~\cite{HEIPRL,PINARD,ATTO}. 
In these experiments, the displacement of the mirror is measured with 
very high sensitivity \cite{HADJAR,ATTO}, and the obtained information
is fed back to the 
mirror via the radiation pressure of another, intensity-modulated, laser
beam, incident on the back of the mirror.
Cold damping is obtained by modulating with the {\em time derivative}
of the homodyne signal, in such a way that
the radiation 
pressure force is proportional to the mirror velocity.
A quantum description of cold damping can be obtained
using either quantum network theory~\cite{GRAS},
or a quantum Langevin description~\cite{LETTER,PRALONG}.
In this latter treatment, cold damping 
implies the following additional term
in the QLE for a generic operator ${\cal O}(t)$,
\begin{equation}\label{DLORO}
\dot{{\cal O}}_{fb}(t)=
\frac{i}{\eta \sqrt{\gamma_{c}}}
\int_{0}^{t} dt'G_{cd}(t')Y_{out}(t-t')
\left[g_{cd}Q(t),{\cal O}(t)\right]\,,
\end{equation}
where $G_{cd}(t)$ and $g_{cd}$ are the corresponding transfer function 
and gain factor. As in the previous case, 
one usually assume a Markovian feedback loop 
with negligible delay. Since one needs a derivative feedback, this 
would ideally imply $G_{cd}(t)= -\delta'(t)$, i.e., 
$\tilde{G}_{cd}(\omega) = i\omega $, $\forall \omega $, even though, 
in practice, it is sufficient to satisfy this 
condition within the detection bandwidth.
In this case, the QLEs for the cold damping 
feedback scheme become
\begin{eqnarray}
\dot{Q}(t)&=&\omega_{m}P(t) \,,
\label{CFBEQ1}\\
\dot{P}(t) &=& -\omega_{m} Q(t) -  
  {\gamma_m} P(t)   +2 G\beta X(t)  
  -g_{cd}\dot{Y}(t)+\frac{g_{cd}}{2\sqrt{\gamma_c \eta}}
  \dot{Y}_{in}^{\eta}(t)
+{\cal W}(t)+f(t)\,,
\label{CFBEQ2}\\
\dot{Y}(t) &=&  -\frac{\gamma_{c}}{2} Y(t)
+2 G \beta Q(t)+ \frac{\sqrt{\gamma_{c}}}{2} Y_{in}(t) \,,
\label{CFBEQ3} \\
\dot{X}(t) &=&  -\frac{\gamma_{c}}{2} X(t)
+ \frac{\sqrt{\gamma_{c}}}{2} X_{in}(t) \,.
\label{CFBEQ4}
\end{eqnarray} 
Adiabatically 
eliminating the cavity mode, and introducing the 
rescaled, dimensionless feedback gain 
$g_{2}=4 G \beta \omega_{m}g_{cd}/\gamma_{m}\gamma_{c}$,
one has
\begin{eqnarray}
&& \dot{Q}(t) = \omega_m P(t), 
 \nonumber \\
&&\dot{P}(t) = -\omega_{m} Q(t) -  
  {\gamma_m} P(t)   +\frac{\sqrt{\gamma_{m}\zeta}}{2}
  X_{in}(t) + {\cal W}(t)  +f(t)- \frac{\gamma_m g_{2}}
 {\omega_m}\dot{Q}(t)-\frac{g_{2}\sqrt{\gamma_{m}}}{\omega_m \sqrt{\zeta}}
 \dot{Y}_{in}(t)
 + \frac{g_{2}\sqrt{\gamma_{m}}}{2\omega_m \sqrt{\eta \zeta}}
 \dot{Y}_{in}^{\eta}(t) \nonumber .
\end{eqnarray}
The presence of an ideal derivative feedback
implies the introduction of two new quantum input noises,
$\dot{Y}_{in}(t)$ and $\dot{Y}_{in}^{\eta}(t)$, whose correlation functions
can be simply obtained by differentiating the corresponding correlation
functions of $Y_{in}(t)$ and $Y_{in}^{\eta}(t)$~\cite{PRALONG}. 
However, as discussed above, these ``differentiated'' correlation 
functions have to be considered as 
approximate expressions valid within the detection bandwidth only.

The two sets of QLE for the mirror Heisenberg operators
show that the two feedback 
schemes are not exactly equivalent. They are however physically 
analogous, as it can be seen, for example, by looking at the 
Fourier transforms of the corresponding mechanical susceptibilities
\cite{LETTER,PRALONG}
\begin{equation}
{\tilde \chi}_{mf}(\omega)=\omega_{m}\left[\omega_{m}^{2}+
	g_{1}\gamma_{m}^{2}-\omega^{2}+i\omega 
	\gamma_{m}\left(1+g_{1}\right)\right]^{-1}
	\label{suscsc}
\end{equation}
for the momentum feedback scheme, and 
\begin{equation}
{\tilde \chi}_{cd}(\omega)=\omega_{m}\left[\omega_{m}^{2}
	-\omega^{2}+i\omega \gamma_{m}\left(1+g_{2}\right)\right]^{-1}
	\label{susccd}
\end{equation}
for cold damping.
These expressions show that in both schemes the main effect of 
feedback is the modification of mechanical damping $\gamma_{m} 
\rightarrow \gamma_{m}(1+g_{i})$ ($i=1,2$). Therefore, also momentum 
feedback provides a cold 
damping effect of increased damping without an increased temperature.
In this latter case, one 
has also a frequency renormalization $\omega_{m}^{2} \rightarrow  
\omega_{m}^{2}+\gamma_{m}^{2}g_{1}$, which is however negligible 
when the mechanical quality factor ${\cal Q}=\omega_{m}/\gamma_{m}$
is large. 

\section{Spectral measurements and their sensitivity}

Spectral measurements are performed whenever the classical force 
$f(t)$ to detect has a characteristic frequency. We adopt a very 
general treatment which can be applied even in the case of 
nonstationary measurements. The explicitly  
measured quantity is the output homodyne photocurrent $Y_{out}(t)$,
and therefore we define the {\em signal} $S(\omega)$ as
\begin{equation}
  	S(\omega)= \left|\int_{-\infty}^{+\infty}dt e^{-i\omega t}\langle 
  	Y_{out}(t)\rangle F_{T_{m}}(t)\right|,
  	\label{signal}
  \end{equation}
where $F_{T_{m}}(t)$ is a ``filter'' function, approximately 
equal to one in the time interval $[0,T_{m}]$ in which the  
measurement is performed, and equal to zero otherwise. The spectral 
measurement is stationary when $T_{m}$ is very large, i.e. is much 
larger than all the typical timescales of the system. Using 
Eq.~(\ref{adiab}) and
the input-output relation (\ref{BOUNDARY}), 
the signal can be rewritten as
\begin{equation}
  	S(\omega)= \frac{8G\beta \eta}{2\pi \sqrt{\gamma_{c}}}
  	\left|\int_{-\infty}^{+\infty}d\omega' {\tilde \chi}(\omega')
  	\tilde{f}(\omega') \tilde{F}_{T_{m}}(\omega -\omega')\right|,
  	\label{signal2}
  \end{equation} 
where $\tilde{f}(\omega)$ and $\tilde{F}_{T_{m}}(\omega )$ 
are the Fourier transforms of the force and 
of the filter function, respectively, and ${\tilde \chi}(\omega)$ is equal to
${\tilde \chi}_{mf}(\omega)$ or ${\tilde \chi}_{cd}(\omega)$, according to the 
feedback scheme considered. 

The noise corresponding to the signal $S(\omega)$ is given by 
its ``variance''; since the signal is zero when $f(t)=0$, the noise
spectrum can be generally written as 
\begin{equation}
  	 N(\omega)= \left\{\int_{-\infty}^{+\infty} dt F_{T_{m}}(t) 
  	\int_{-\infty}^{+\infty} dt' F_{T_{m}}(t') e^{-i\omega (t-t')} 
  	 \langle Y_{out}(t)Y_{out}(t')\rangle_{f=0} \right\}^{1/2},
  	\label{noise}
  \end{equation}
where the subscript $f=0$ means evaluation
in the absence of the external force. Using again 
(\ref{adiab}), Eqs.~(\ref{BOUNDARY}), 
and the input noises correlation functions (\ref{INCOR1})-(\ref{INCOR2}) and 
(\ref{INCORETA1})-(\ref{INCORETA3}), the 
spectral noise can be rewritten as
\begin{equation}
 N(\omega)= \left\{ \frac{(8G\beta \eta)^{2}}{\gamma_{c}}
  	\int_{-\infty}^{+\infty}dt F_{T_{m}}(t) 
  	\int_{-\infty}^{+\infty}dt' F_{T_{m}}(t') e^{-i\omega 
  	(t-t')}C(t,t')+\eta \int_{-\infty}^{+\infty}dt F_{T_{m}}(t)^{2} 
  	\right\}^{1/2},
  	\label{noise2}
  \end{equation}
where $C(t,t')=\langle Q(t)Q(t')+Q(t')Q(t)\rangle/2 $ is the 
symmetrized correlation function of the oscillator position.
This very general expression of the noise spectrum is nonstationary
because it depends upon the nonstationary correlation function 
$C(t,t')$. The last term in Eq.~(\ref{noise2})
is the shot noise term due to the radiation input noise.

\subsection{Stationary spectral measurements}

Spectral measurements are usually performed in the stationary case, 
that is, using a measurement time $T_{m}$ much 
larger than the typical oscillator timescales. The most significant
timescale is the mechanical
relaxation time, which is $\gamma_{m}^{-1}$ in the absence of feedback 
and $[\gamma_{m}(1+g_{i})]^{-1}$ ($i=1,2$) in the presence of 
feedback. In the stationary case,
the oscillator is relaxed to equilibrium and, redefining $t'=t+\tau$,
the correlation function $C(t,t')=C(t,t+\tau)$ in Eq.~(\ref{noise2}) 
is replaced by the {\em stationary} 
correlation function $C_{st}(\tau)=\lim_{t\to \infty} C(t,t+\tau)$. 
Moreover, for very large $T_m$,
one has $F_{T_{m}}(t+\tau) \simeq F_{T_{m}}(t)\simeq 1$ 
and, defining the 
measurement time $T_{m}$ so that $T_{m}=\int dt F_{T_{m}}(t)^{2}$,  
Eq.~(\ref{noise2}) assumes the form
\begin{equation}
  	N(\omega)= \left\{ \left[\frac{(8G\beta \eta)^{2}}{\gamma_{c}}
  	N_{Q}^{2}(\omega)+\eta\right]T_{m}  
  	\right\}^{1/2},
  	\label{noisesta}
  \end{equation} 
where 
\begin{equation}
N_{Q}^{2}(\omega)=\int_{-\infty}^{+\infty} d\tau e^{-i\omega \tau}C(\tau) ,
\label{noiseq}
\end{equation}
is the stationary position noise spectrum. 
This noise spectrum can be evaluated by solving the quantum Langevin 
equations in the presence of the two feedback schemes and Fourier 
transforming ~\cite{PRALONG}. One obtains 
\begin{equation}
N_{Q,mf}^{2}(\omega)=\gamma_{m} \left 
|{\tilde \chi}_{sc}(\omega)\right|^{2}\left[\frac{\zeta }{4}
+\frac{g_1^2}{4 \eta \zeta }
\frac{\omega^2+\gamma_m^2}{\omega_m^2}
|\tilde{G}_{mf}(\omega)|^{2}+\frac{\omega}{2 \omega_{m}} \coth
\left(\frac{\hbar \omega}{2 k_{B} T}\right)\Theta_{[-\varpi,\varpi]}(\omega) 
\right].
\label{qspesc}
\end{equation}
for the momentum feedback scheme and
\begin{equation}
N_{Q,cd}^{2}(\omega)=\gamma_{m}\left 
|{\tilde \chi}_{cd}(\omega)\right|^{2}\left[\frac{\zeta}{4}
+\frac{g_2^2}{4 \eta \zeta }
\frac{|\tilde{G}_{cd}(\omega)|^{2}}{\omega_m^2}
+\frac{\omega}{2 \omega_{m}} \coth
\left(\frac{\hbar \omega}{2 k_{B} T}\right)\Theta_{[-\varpi,\varpi]}(\omega) 
\right],
\label{qspecd}
\end{equation}
for the cold damping scheme, where $\tilde{G}_{i}(\omega)$ (i=mf,cd) 
are the Fourier transforms of the feedback transfer functions and 
$\Theta_{I}(\omega)$ is a gate function equal to $1$ within the 
frequency interval $I$ and zero outside.
The position noise spectrum for the momentum feedback essentially
coincides with that already obtained in \cite{MVTPRL}, 
except that in that paper the high-temperature ($\coth(\hbar \omega
/2k_B T) \simeq 2 k_B T/\hbar \omega$), Markovian feedback 
($\tilde{G}_{mf}(\omega)\simeq 1$), and infinite cutoff ($\varpi \to 
\infty$) approximations have been considered. 
The noise spectrum in the cold damping case of Eq.~(\ref{qspecd})
instead essentially reproduces the one obtained in \cite{COURTY}, with the
difference that in Ref.~\cite{COURTY} the homodyne detection
efficiency $\eta$ is set equal to one, and the Markovian feedback 
($\tilde{G}_{cd}(\omega)\simeq i\omega $), and infinite cutoff ($\varpi \to 
\infty$) approximations have been again considered. The comparison
between Eqs.~(\ref{qspecd}) and (\ref{qspesc}) shows once again the
similarities of the two schemes. The only differences lie in the 
different susceptibilities and in the feedback-induced noise term,
which has an additional $\gamma_m^2/\omega_m^2$ factor in the momentum 
feedback case, which is however usually negligible with good mechanical quality
factors. In fact, the two noise spectra 
are practically indistinguishable in a very large parameter region.

The effectively detected position noise spectrum is not given
by Eqs.~(\ref{qspecd}) and (\ref{qspesc}), but it is the noise 
spectrum associated to the output homodyne photocurrent of Eq.~(\ref{noisesta}) 
rescaled to a position spectrum.
This homodyne-detected position noise spectrum is actually subject 
also to cavity filtering, yielding an experimental high frequency 
cutoff $\gamma_{c}$, which however does not appear in 
our expressions because we have adiabatically eliminated the 
cavity mode from the beginning. Therefore the noise spectrum 
derived here is correct only for $\omega < \gamma_{c}$. However,
this is not a problem because the frequencies of interest (i.e. those 
within the detection bandwidth) are always much smaller than 
$\gamma_{c}$, and also than $\varpi$.  Within the detection bandwidth
one can safely approximate $\tilde{G}_{mf}(\omega)\simeq 1$, 
$\tilde{G}_{cd}(\omega)\simeq i\omega $ and 
$\Theta_{[-\varpi,\varpi]}(\omega)=1$ in 
Eqs.~(\ref{qspecd}) and (\ref{qspesc}). Therefore, using 
Eq.~(\ref{noisesta}), the detected position noise spectrum can be 
written
\begin{equation}
N_{Q,det}^{2}(\omega)=\gamma_{m}\left 
|{\tilde \chi}_{i}(\omega)\right|^{2}\left[\frac{\zeta}{4}
+\frac{g_i^2}{4 \eta \zeta}
\frac{\omega^2+\delta_{i,1}\gamma_m^2}{\omega_m^2}
+\frac{\omega}{2 \omega_{m}} \coth
\left(\frac{\hbar \omega}{2 k_{B} T}\right)
\right]+\frac{1}{4 \eta \zeta \gamma_{m}},
\label{qspedet}
\end{equation}
where $i=1$ refers to the momentum feedback case and 
$i=2$ to the cold damping case.

This spectrum has four contributions: the radiation pressure noise 
term, proportional to the input power $\wp$, 
the feedback-induced term proportional to the 
squared gain and inversely proportional to $\wp$, the Brownian
motion term which is independent of $\wp$, and the shot noise
term inversely proportional to $\wp$. The main effect of feedback 
on the noise spectrum is the
modification of the susceptibility due to the increase of damping,
yielding the suppression and widening of the resonance peak.
This peak suppression in the noise spectrum
has been already predicted and illustrated in
\cite{MVTPRL,COURTY}, and experimentally verified for the cold damping
case in \cite{HEIPRL,PINARD}. 
It has been shown~\cite{COURTY,PRALONG}
that both feedback schemes are able to 
arbitrarily reduce the displacement noise at resonance. 
This noise reduction at resonance is similar to that
occurring to an oscillator with increasing damping, except that in 
the present case, also
the feedback-induced noise increases with the gain, and it
can be kept small only if the input power is correspondingly increased
in order to maintain the optimal value minimizing the 
noise~\cite{PRALONG}. This arbitrary reduction of the position noise
in a given frequency bandwidth with increasing feedback gain does not 
hold if the input power $\zeta$ is kept fixed. In this latter
case, the noise has a 
frequency-dependent lower bound which cannot be overcome by 
increasing the gain.

\begin{figure}
   \begin{center}
   \begin{tabular}{c}
   \includegraphics[height=5cm]{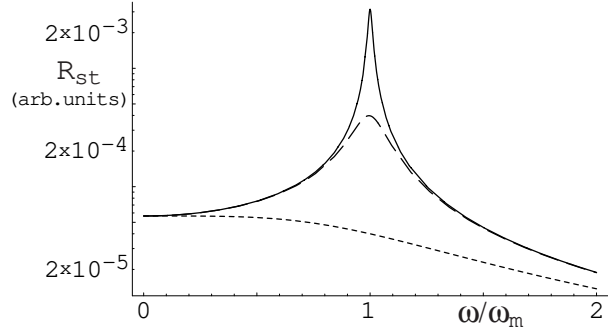}
   \end{tabular}
   \end{center}
   \caption[snrsta] 
   { \label{snrsta} 
   Stationary SNR as a function of frequency in the case of an ideal 
impulsive force, i.e., ${\tilde f}(\omega)=$ const. The full line 
refers to the case with no feedback, the dashed line to the case 
with $g_{1}=g_{2}=10^{4}$, and the dotted line to the case with
$g_{1}=g_{2}=10^{5}$ (the two feedback schemes give 
indistinguishable results in these cases). The other parameters are 
${\cal Q}=10^{5}$, $\zeta =10$, $k_{B}T/\hbar \omega_{m}=
10^{5}$, and $\eta =0.8$. At a given frequency, the stationary 
SNR decreases for increasing feedback gain.}
\end{figure}

In the case of stationary spectral measurements also the
expression of the signal simplifies. In fact, one has
$ \tilde{F}_{T_{m}}(\omega) \simeq \delta(\omega)$,
and Eq.~(\ref{signal2}) becomes
$
  	S(\omega)= 8G\beta \eta| {\tilde \chi}(\omega)
  	\tilde{f}(\omega) |/2\pi \sqrt{\gamma_{c}}$.
The stationary SNR, ${\cal R}_{st}(\omega)$, is now simply obtained 
dividing this signal by the noise of 
Eq.~(\ref{noisesta}), 
\begin{equation}
 {\cal R}_{st}(\omega)= 
  	|\tilde{f}(\omega)|\left\{\gamma_{m}T_{m}\left[\frac{
  	\omega}{2\omega_{m}}\coth\left(\frac{\hbar \omega}{2k_{B}T}\right)+
  	\frac{\zeta}{4} 
   +\frac{1}{4 \eta \zeta }\left(\frac{g_{i}^{2}}{\omega_{m}^{2}} 
   \left(\omega^{2}+\delta_{i,1}\gamma_{m}^{2}\right)
  +\frac{1}{\gamma_{m}^{2}\left|{\tilde \chi}_{i}(\omega)\right|^{2}}\right)
  \right]\right\}^{-1/2},
  	\label{snr}
  \end{equation}  
where again $i=1$ refers to the momentum feedback case and $i=2$ to 
the cold damping case.
It is easy to see that, in both cases,
feedback {\em always lowers} the stationary SNR at any frequency,
(except at $\omega=0$, where the SNR for the cold damping case does not 
depend upon the feedback gain). This is shown in Fig.~\ref{snrsta}, 
where the stationary SNR in the case of an ideal impulsive force (that 
is, ${\tilde f}(\omega)$ is a constant) is plotted for three values of 
the feedback gain. The curves refer to both feedback schemes because
the two cases $i=1,2$ gives always practically indistinguishable results, 
except for very low values of ${\cal Q}$.
This result can be easily explained.
In fact, the main effect of 
feedback is to decrease 
the mechanical susceptibility at resonance 
(see Eqs.~(\ref{suscsc}) and (\ref{susccd})), so that the oscillator is 
less sensitive not only to the noise but also to the signal.
Therefore, even though the two feedback schemes are able to provide 
efficient cooling and noise reduction in narrow bandwidths for
the mechanical mode, they cannot be used to improve the sensitivity
of the optomechanical device for stationary measurements. In the next 
section we shall see how cooling via feedback can be used to improve
the sensitivity for the detection of impulsive forces, using an 
appropriate nonstationary strategy.

\section{High-sensitive nonstationary measurements}

The two feedback schemes discussed here achieve noise reduction 
through a modification of the mechanical susceptibility. However, this
modification does not translate into a sensitivity improvement because at the
same time it strongly degrades the detection of the signal. 
The sensitivity of position measurements would be improved if 
the oscillator mode could keep its intrinsic susceptibility, unmodified by
feedback, together with the reduced noise achieved by the feedback loop.
This is obviously impossible in stationary conditions, but a
situation very similar to this ideal one can be realized in the case of
the detection of an {\em impulsive} force, that is, with a 
time duration 
$\sigma$ much shorter than the mechanical relaxation time (in the absence
of feedback), $\sigma \ll 1/\gamma_{m}$. 
In fact, one could use the following nonstationary strategy: prepare
at $t=0$ the mirror mode in the stationary state cooled by feedback,
then suddenly turn off the feedback loop and perform the spectral 
measurement in the presence of the impulsive force
for a time $T_m $, such that $\sigma \ll T_m  \ll 1/\gamma_m$.
In such a way, the force spectrum is still
well reproduced, and the mechanical susceptibility is the one without feedback
(even though modified by the short measurement time $T_m \ll 1/\gamma_m$).
At the same time, the mechanical mode is far from equilibrium 
during the whole 
measurement, and its noise spectrum is different from the 
stationary form of Eq.~(\ref{qspedet}), being mostly determined by the 
{\em cooled} initial state. As long as $T_m \ll \gamma_m$, heating, that is,
the approach to the hotter equilibrium without feedback, will not
affect and increase too much the noise spectrum.
Therefore, one expects 
that as long as the measurement time is sufficiently short,
the SNR for the detection of the impulsive 
force (which has now to be evaluated using the most general expressions 
(\ref{signal2}) and (\ref{noise2})) can be significantly increased by 
this nonstationary strategy.

It is instructive to evaluate explicitely the nonstationary
noise spectrum of Eq.~(\ref{noise2}) 
for the above measurement strategy.
Simple analytical results are obtained by choosing
the following filter function $
F_{T_m}(t) =\theta(t) e^{-t/2T_m}$
($\theta(t)$ is the Heavyside step function),
satisfying $\int dt F_{T_m}(t)^2 = T_m$. Let us consider the cold 
damping case first. Solving the QLE of the system
by taking the equilibrium state in the presence of feedback as initial 
condition, one arrives at the following expression for the detected 
nonstationary noise spectrum~\cite{PRALONG}
\begin{equation}
 N_{Q,non}^{2}(\omega)= \gamma_{m}
 \left|{\tilde \chi_0}(\omega-i/2T_m)\right|^2 \left[
 \frac{\omega^2+(1/2T_m+\gamma_m)^2}{\omega_m^2 \gamma_{m}T_{m}}  
 \langle Q^2\rangle_{st} +
\frac{\langle P^2\rangle_{st}}{\gamma_{m}T_{m}}+ \left( \frac{\zeta}{4}+
\frac{k_B T}{\hbar \omega_{m}}\right)\right] +\frac{1}{4\eta \zeta 
\gamma_{m}} ,
  	\label{noisenosta2}
  \end{equation}
where $\tilde {\chi}_0$ is the Fourier transform of the 
susceptibility in the absence of feedback (see Eq.~(\ref{suscsc}) 
with $g_{1}=0$ or Eq.~(\ref{susccd}) 
with $g_{2}=0$) and we have used the 
high temperature approximation $\coth(\hbar 
\omega /2k_B T) \simeq 2 k_B T/\hbar \omega$ for the Brownian noise. 
Moreover, $\langle Q^2\rangle_{st}$ and $\langle P^2\rangle_{st}$ are the 
stationary variances in the presence of feedback, 
\begin{eqnarray}
&& \langle Q^2 \rangle_{st} = \lim_{t \to \infty}\langle Q^2(t) \rangle
= C_{st}(0) = \int_{-\infty}^{+\infty} \frac{d \omega }{2 
\pi}N_{Q^{2},cd}(\omega)
\label{perspe1} \\
&& \langle P^2 \rangle_{st} = \lim_{t \to \infty}\langle P^2(t) \rangle
 = \int_{-\infty}^{+\infty} \frac{d \omega }{2 
 \pi}\frac{\omega^{2}}{\omega_{m}^{2}}N_{Q^{2},cd}(\omega),
\label{perspe2}
\end{eqnarray}
where $N_{Q^{2},cd}(\omega)$ is given by Eq.~(\ref{qspecd}).
The nonstationary noise spectrum for the momentum feedback case is 
analogous to that of Eq.~(\ref{noisenosta2}), except that one has to 
use the corresponding stationary values as initial conditions (there 
is an additional term due to the fact that $\langle 
QP+PQ\rangle_{st}\neq 0$ for momentum feedback~\cite{PRALONG}). 
However, as for the stationary case, one can check that
the two feedback schemes give 
indistinguishable results in a large parameter region. Therefore 
we shall discuss the cold damping case only from now on, even though 
the same results also apply to the momentum feedback case with the 
replacement $g_{2} \to g_{1}$.

\begin{figure}
   \begin{center}
   \begin{tabular}{c}
   \includegraphics[height=5cm]{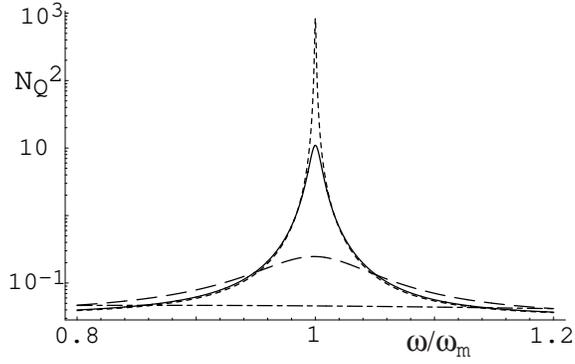}
   \end{tabular}
   \end{center}
   \caption[nonstatm] 
   { \label{nonstatm} 
Nonstationary noise spectrum 
for different values of the measurement time,  
$\gamma_m T_m =10^{-1}$ (dotted line),
$\gamma_m T_m =10^{-2}$ (full line), $\gamma_m T_m =10^{-3}$ (dashed line),
$\gamma_m T_m =10^{-4}$ (dot-dashed line). 
The figure refers to the cold damping feedback scheme, but the curves 
are indistinguishable from that obtained with the momentum feedback,
using the same parameters, ${\cal Q}=10^4$, $\zeta =10$, $g_1=g_2=10^3$,
$k_BT/\hbar \omega_m = 10^5$, $\eta =0.8$.}
\end{figure}

It is easy to check from Eq.~(\ref{noisenosta2}) 
that the stationary noise spectrum corresponding 
to the situation with no feedback is recovered in the limit of large 
$T_m$, as expected, when the terms inversely proportional to 
$\gamma_{m}T_m$ and depending on the initial conditions
become negligible, and 
${\tilde \chi_0}(\omega-i/2T_m) \to {\tilde \chi_0}(\omega)$. In the opposite 
limit of small $T_m$ instead, 
the terms associated to the {\em cooled}, initial conditions 
are dominant, and since the stationary terms are still
small, this means having a reduced, nonstationary noise spectrum.
This is clearly visible in Fig.~\ref{nonstatm}, where the nonstationary
noise spectrum 
is plotted for different values of the
measurement time $T_m$, $\gamma_m T_m =10^{-1}$ (dotted line),
$\gamma_m T_m =10^{-2}$ (full line), $\gamma_m T_m =10^{-3}$ (dashed line),
$\gamma_m T_m =10^{-4}$ (dot-dashed line). The resonance peak is 
significantly suppressed for decreasing $T_m$, 
even if it is simultaneously widened, so that
one can even have a slight increase of noise out of resonance.

\begin{figure}
   \begin{center}
   \begin{tabular}{c}
   \includegraphics[width=16cm]{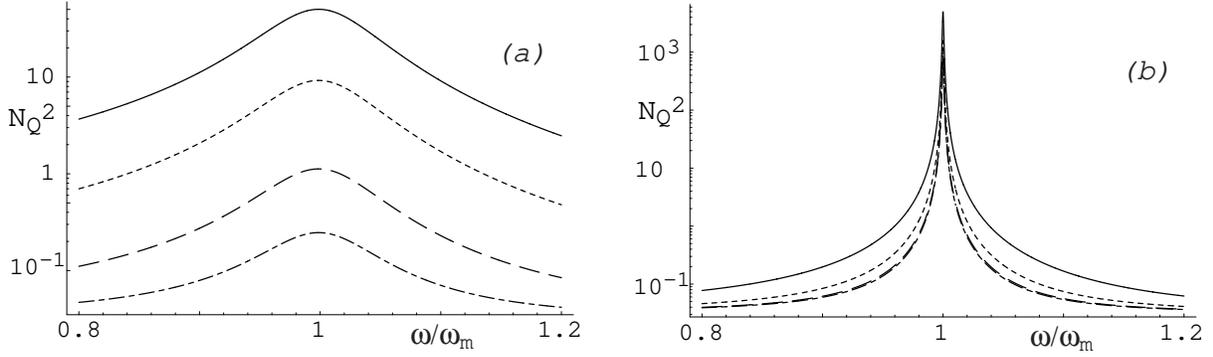}
   \end{tabular}
   \end{center}
   \caption[nonstag] 
   { \label{nonstag} 
Nonstationary noise spectrum 
for different values of the feedback gain,
$g_2=1$ (full line), $g_2=10$ (dotted line), $g_2=10^2$ (dashed),
$g_2=10^3$ (dot-dashed), with fixed measurement time,
$\gamma_m T_m=10^{-3}$ (a), and $\gamma_m T_m=10^{-1}$ (b). 
(a) corresponds to a strongly nonstationary
condition, in which the noise is significantly suppressed,
thanks to the cooled initial condition.
In (b) the stationary terms becomes important and the noise
reduction due to feedback cooling is less significant.
The other parameters are ${\cal Q}=10^4$, $\zeta =10$,
$k_BT/\hbar \omega_m = 10^5$, $\eta =0.8$.}
\end{figure}

The effect of the terms depending upon the feedback-cooled initial 
conditions on the nonstationary noise is shown in 
Fig.~\ref{nonstag}, where the noise spectrum
is plotted for different values of the feedback gain at a fixed value
of $T_m$. In Fig.~\ref{nonstag}a, $N_{Q}^{2}(\omega)$ is plotted at 
$\gamma_m T_m=
10^{-3}$ for $g_2=1$ (full line), $g_2=10$ (dotted line), $g_2=10^2$ (dashed),
$g_2=10^3$ (dot-dashed). For this low value of $\gamma_m T_m$, 
the noise terms depending on the
initial conditions are dominant, and increasing the feedback gain implies
reducing the initial variances, and therefore an approximately 
uniform noise suppression at all frequencies.
In Fig.~\ref{nonstag}b, $N_{Q}^{2}(\omega)$ is instead plotted 
at $\gamma_m T_m=
10^{-1}$ for $g_2=1$ (full line), $g_2=10$ (dotted line), $g_2=10^2$ (dashed),
$g_2=10^3$ (dot-dashed). In this case, the feedback-gain-independent,
stationary terms become important,
and the effect of feedback on the noise spectrum
becomes negligible. 

\begin{figure}[ht]
   \begin{center}
   \begin{tabular}{c}
   \includegraphics[width=16cm]{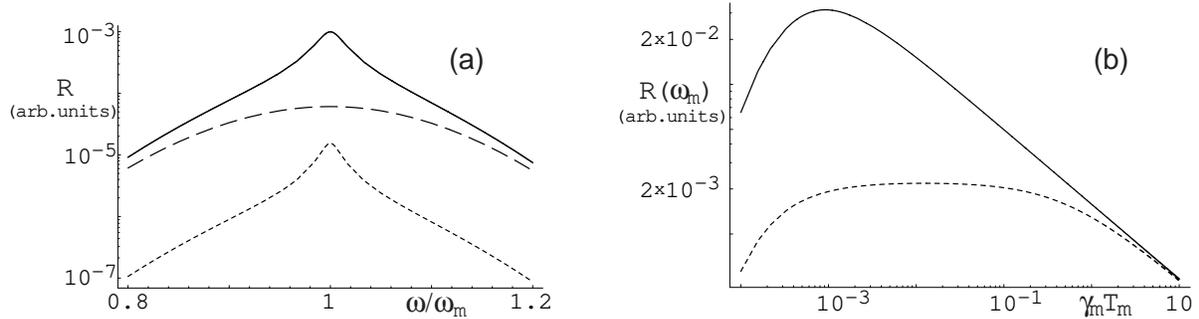}
   \end{tabular}
   \end{center}
   \caption[snrnontot] 
   { \label{snrnontot} 
(a) Spectrum of the nonstationary SNR, ${\cal R}(\omega)$, with and without 
feedback cooling of the initial state. The full line refers to a 
nonstationary measurement, $\gamma_{m}T_{m}=10^{-3}$,
in the presence of feedback, $g=2\cdot 10^{3}$ (the two feedback 
schemes give indistinguishable curves); the dashed line refers to the 
no-feedback case, and with the same, short, measurement time 
$\gamma_{m}T_{m}=10^{-3}$. Finally, the dotted line refers to a ``standard 
measurement'', without feedback, and in the stationary limit
$\gamma_{m}T_{m}=10$. (b) Nonstationary SNR at resonance, 
${\cal R}(\omega_{m})$, with and without 
feedback cooling of the initial state, plotted as a function of the 
rescaled measurement time $\gamma_{m}T_{m}$.
The full line refers to the case with feedback-cooled initial
conditions ($g=2\cdot 10^{3}$). The dotted line refers to the 
no-feedback case, $g=0$. 
The other parameters are
$\omega_{f}=\omega_{m}$, $ \gamma_{m}\sigma = 10^{-4}$, 
$\gamma_{m}t_{1}=3 \cdot 10^{-4}$, 
${\cal Q}=10^{5}$, $\zeta=10$, $\eta =0.8$, $k_{B}T/\hbar 
\omega_{m}=10^{5}$.}
\end{figure}

The significant noise reduction attainable at short measurement times
$\gamma_{m}T_{m} \ll 1$ is not only due to the feedback-cooled 
initial conditions, but it is also caused by the effective 
reduction of the mechanical susceptibility given by the short 
measurement time, ${\tilde \chi}_{0}(\omega) 
\to {\tilde \chi}_{0}(\omega-i/2T_{m})$. 
This lowered susceptibility yields a simultaneous 
reduction of the signal at small measurement times $\gamma_{m}T_{m} 
\ll 1$, and therefore the behavior of the nonstationary SNR 
may be nontrivial. However, one expects that impulsive forces at least
can be satisfactorily 
detected using a short measurement time, because the noise can be kept 
very small and the corresponding sensitivity increased.
Let us check this fact considering the case of 
the impulsive force
\begin{equation}
f(t)=f_{0}\exp\left[-(t-t_{1})^{2}/2\sigma^{2}\right]
  	\cos\left(\omega_{f}t\right),
\label{force}
\end{equation}
where $\sigma$ is the force duration, $t_{1}$ its ``arrival time'', and
$\omega_{f}$ its carrier frequency. The corresponding SNR is obtained 
dividing the signal of Eq.~(\ref{signal}) by the nonstationary noise
spectra of Eq.~(\ref{noisenosta2}), and it 
is shown in Fig.~\ref{snrnontot}. As anticipated, the 
sensitivity of the optomechanical device is improved using
feedback in a nonstationary way. In Fig.~\ref{snrnontot}a, the spectral SNR, 
${\cal R}(\omega)$, is plotted for different values of feedback gain 
and measurement time.
The full line refers to $g_1=g_{2}=g=2\cdot 10^{3}$ and 
$\gamma_{m}T_{m}=10^{-3}$, the dashed line to the situation with 
no feedback and the same measurement time, $g=0$ and 
$\gamma_{m}T_{m}=10^{-3}$; finally the dotted line refers to a 
``standard'' measurement, that is, no feedback and a stationary 
measurement, with a long measurement time, $\gamma_{m}T_{m}=10$. 
The proposed nonstationary measurement scheme,
``cool and measure'', gives the highest sensitivity. This is 
confirmed also by Fig.~\ref{snrnontot}b, where the SNR at resonance,
${\cal R}(\omega_{m})$, when feedback cooling is used with
$g=2\cdot 10^{3}$ (full line), and without feedback cooling
(dotted line), is plotted as a function of the rescaled
measurement time $\gamma_{m}T_{m}$. The preparation of the mirror in 
the cooled initial state yields a better sensitivity for any 
measurement time. As expected, the SNR in the presence of 
feedback approaches that without feedback in the stationary limit
$\gamma_{m}T_{m} \gg 1$, when the effect of the initial cooling 
becomes irrelevant.
Fig.~\ref{snrnontot} refer 
to a resonant ($\omega_{f}=\omega_{m}$)
impulsive force with $ \gamma_{m}\sigma = 10^{-4}$ and 
$\gamma_{m}t_{1}=3 \cdot 10^{-4}$, while the other parameters are
${\cal Q}=10^{5}$, $\zeta=10$, $\eta =0.8$, $k_{B}T/\hbar 
\omega_{m}=10^{5}$.

The proposed nonstationary strategy 
can be straightforwardly applied whenever the ``arrival 
time'' $t_{1}$ of the impulsive force
is known: feedback has to be 
turned off just before the arrival of the force. However, the 
scheme can be easily adapted also to the case of an impulsive force with an 
{\em unknown arrival time}, as for example, that of
a gravitational wave passing through an interferometer. In this case 
it is convenient to repeat the process many times, i.e., 
subject the oscillator to cooling-heating cycles. 
Feedback is turned off for a time $T_{m}$ during which the spectral measurement 
is performed and the oscillator starts heating up. Then feedback is 
turned on and the oscillator is cooled, and then the process is 
iterated. This cyclic cooling strategy improves the
sensitivity of gravitational wave detection 
provided that the cooling time $T_{cool}$,
which is of the order of $1/\left[\gamma_{m}(1+g_{i})\right]$, is much 
smaller than $T_{m}$, which is verified at sufficiently large gains.
Cyclic cooling has been proposed, in a qualitative way, 
to cool the violin modes of 
a gravitational waves interferometer in \cite{PINARD}, and its 
capability of improving the high-sensitive detection of impulsive forces
has been first shown in \cite{LETTER}. 
In the case of a random, uniformly distributed, arrival time
$t_1$ and in the impulsive limit $\sigma \ll T_{m}$, 
the performance of the cyclic cooling scheme is well characterized by 
a time averaged SNR, i.e., 
\begin{equation}
\langle {\cal R}(\omega)\rangle = 
\frac{1}{T_{m}+T_{cool}}\left\{\int_{0}^{T_{m}}dt_{1}
{\cal R}(\omega,t_{1})+\int_{T_{m}}^{T_{m}+T_{cool}}dt_{1}
{\cal R}(\omega,t_{1})_{cool}\right\},
	\label{snrmedio}
\end{equation}
where ${\cal R}(\omega,t_{1})$ is the nonstationary SNR at a given 
force arrival time $t_{1}$ discussed in this section, and ${\cal 
R}(\omega,t_{1})_{cool}$ is the nonstationary SNR one has during the 
cooling cycle, which means with feedback turned on and with uncooled 
initial conditions. It is easy to understand that ${\cal 
R}(\omega,t_{1})_{cool} \ll {\cal 
R}(\omega,t_{1})$, and, since it is also $T_{cool} \ll T_{m}$, 
the second term in Eq.~{\ref{snrmedio}) can be neglected, so that
\cite{LETTER},
\begin{equation}
\langle {\cal R}(\omega)\rangle \simeq 
\frac{1}{T_{m}+T_{cool}}\int_{0}^{T_{m}}dt_{1}
{\cal R}(\omega,t_{1}).
	\label{snrmedio2}
\end{equation}
 
This time-averaged SNR can be significantly improved 
by cyclic cooling, as it is shown in Fig.~\ref{snrme2}, 
where $\langle {\cal R}(\omega)\rangle$ is plotted both with and 
without feedback. The full line describes 
the time-averaged SNR subject to cyclic feedback-cooling 
with $g=2\cdot 10^{3}$, $\gamma_{m}T_{m}=10^{-3}$, and
$T_{cool}=10^{-3}T_{m}$. In the absence of feedback, in the case of 
an impulsive 
force with unknown arrival time and duration $\sigma$, 
the best strategy is to perform repeated measurements of 
duration $T_{m}$ without any cooling stage. The measurement time 
$T_{m}$ can be optimized considering that it has to be longer
than $\sigma$, and at the same time it has not to be too long, in 
order to have a good SNR (see the dotted line in 
Fig.~\ref{snrnontot})b.
In this case, the time-averaged SNR can be written as
\begin{equation}
\langle {\cal R}_{0}(\omega)\rangle \simeq 
\frac{1}{T_{m}}\int_{0}^{T_{m}}dt_{1}
{\cal R}_{0}(\omega,t_{1}),
	\label{snrmedio3}
\end{equation}
where ${\cal R}_{0}(\omega,t_{1})$ is the SNR evaluated for $g=0$.
The dashed line in Fig.~\ref{snrme2} refers to this case without 
feedback, and with $\gamma_{m}T_{m}=10^{-3}$. 
The other parameter values
are the same as in Fig.~\ref{snrnontot} and in this 
case, cyclic cooling provides
an improvement at resonance by a factor $16$ with respect to 
the case with no feedback.
As suggested in Ref.~\cite{PINARD},
one could use nonstationary cyclic feedback to cool the violin modes 
in gravitational-wave interferometers, which have sharp resonances 
within the detection band. One expects that single gravitational
bursts, having a duration smaller than the cooling cycle period, 
could be detected in this way.

\begin{figure}[ht]
   \begin{center}
   \begin{tabular}{c}
   \includegraphics[height=5cm]{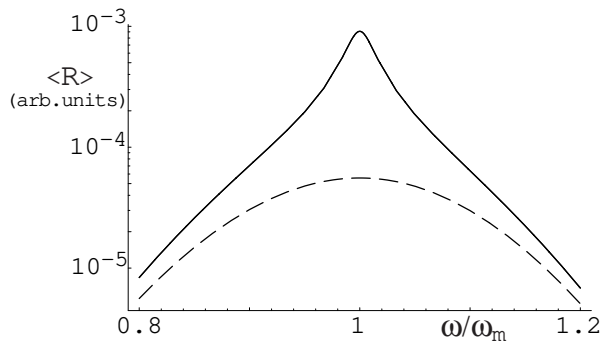}
   \end{tabular}
   \end{center}
   \caption[snrme2] 
   { \label{snrme2} 
Time averaged spectral SNR 
with and without cyclic cooling. The full line refers to 
cyclic cooling with $\gamma_{m}T_{m}=10^{-3}$,
$g=2\cdot 10^{3}$, and $T_{cool}=10^{-3}T_{m}$ (the two feedback 
schemes give indistinguishable curves). The dashed line refers to the 
no-feedback case, with the same measurement time 
$\gamma_{m}T_{m}=10^{-3}$ (see Eq.~(\protect\ref{snrmedio3})).
The other parameters are
$\omega_{f}=\omega_{m}$, $ \gamma_{m}\sigma = 10^{-4}$,  
${\cal Q}=10^{5}$, $\zeta=10$, $\eta =0.8$, $k_{B}T/\hbar 
\omega_{m}=10^{5}$.}
\end{figure}

\end{document}